\documentclass[conference, a4paper]{IEEEtran}

\IEEEoverridecommandlockouts
\IEEEpubid{\makebox[\columnwidth]{\textbf{979-8-3315-6655-5/25/\$31.00 ©2025 IEEE}\hfill}
\hspace{\columnsep}\makebox[\columnwidth]{}}
\usepackage[turkish]{babel}
\usepackage[utf8]{inputenc}
\usepackage[T1]{fontenc}
\usepackage{cite}

\usepackage[cmex10]{amsmath}
\usepackage{multirow}
\usepackage{array}
\usepackage[lofdepth,lotdepth]{subfig}
\usepackage{graphicx}

\AtBeginDocument{%
  \renewcommand{\tablename}{TABLO}
  \renewcommand{\refname}{Kaynaklar}
}

\AtBeginDocument{%
  \renewcommand\abstractname{Abstract}
}

\begin{document}


\title{TurkEmbed4Retrieval: Türkçe İçin Geri Getirme Görevine Özel Gömme Modeli\\
TurkEmbed4Retrieval: Turkish Embedding Model for Retrieval Task
\thanks{\textit{\underline{Citation}}: 
\textbf{Özay Ezerceli, Gizem Gümüşçekiçci, Tuğba Erkoç, Berke Özenç. "TurkEmbed4Retrieval: Turkish Embedding Model for Retrieval Task." 2025 33rd Signal Processing and Communications Applications Conference (SIU), 2025. DOI: 10.1109/SIU66497.2025.11112338}} 
}

\author {
    \IEEEauthorblockN {
        Özay Ezerceli
    }
    \IEEEauthorblockA {
        Newmind AI \\
        Istanbul, Türkiye \\
        \texttt{oezerceli@newmind.ai}
    }
    \and
    \IEEEauthorblockN {
        Gizem Gümüşçekiçci, Tuğba Erkoç, Berke Özenç
    }
    \IEEEauthorblockA {
        Işık Üniversitesi \\
        Istanbul, Türkiye \\
        \texttt{\{gizem.gumuscekicci,tugba.erkoc,berke.ozenc\}@isikun.edu.tr}
    }
}


%

\maketitle

\begin{ozet}
Bu çalışmada, öncelikle Doğal Dil Çıkarımı (DDÇ) ve Anlamsal Metin Benzerliği (AMB) görevleri için geliştirilen TurkEmbed modelinin, MS-Marco-TR veri seti üzerinde ince ayar yapılarak geri getirme görevlerine uygun hale getirilmesini sağlayan TurkEmbed4Retrieval modelini tanıtıyoruz. Model, Matruşka temsili öğrenme ve özel tasarlanmış negatif çiftlerin sıralanması kayıp fonksiyonu gibi ileri seviye eğitim teknikleri kullanılarak optimize edilmiştir. Yapılan kapsamlı deneyler, TurkEmbed4Retrieval'ın, geri getirme metriklerinde Turkish-colBERT modelini Scifact-TR veri kümesinde \%19–26 oranında geçtiğini göstermektedir. Bu bağlamda, modelimiz, Türkçe bilgi getirme sistemleri için yeni bir çıtaya ulaşmaktadır.
\end{ozet}
\begin{IEEEanahtar}
türkçe gömme modeli,bilgi getirme, ms-marco-tr, doğal dil işleme, anlamsal benzerlik
\end{IEEEanahtar}

\begin{abstract}
In this work, we introduce TurkEmbed4Retrieval, a retrieval-specialized variant of the TurkEmbed model originally designed for Natural Language Inference (NLI) and Semantic Textual Similarity (STS) tasks. By fine-tuning the base model on the MS‐MARCO-TR dataset using advanced training techniques, including Matryoshka representation learning and a tailored multiple negatives ranking loss, we achieve state-of-the-art (SOTA) performance for Turkish retrieval tasks. Extensive experiments demonstrate that our model outperforms Turkish‐colBERT by 19–26\% on key retrieval metrics for the Scifact-TR dataset, thereby establishing a new benchmark for Turkish information retrieval.
\end{abstract}



%
\IEEEpeerreviewmaketitle

\IEEEpubidadjcol

\section{G{\footnotesize İ}r{\footnotesize İ}ş}

Hesaplamalı Dilbilimin bir kolu olan Doğal Dil İşleme (DDİ) makinelerin insan dilini anlaması, işlemesi ve insan dili kullanarak üretim yapmasına odaklanır ve insan dilinin karmaşıklıklarını anlama ve işleme yeteneği sayesinde modern teknolojinin temel taşlarından biri haline gelmiştir. DDİ, makine çevirisi, metin özetleme, metinden duygu analizi ve iğneleme tespiti gibi birçok problemde kullanılır \cite{ezerceli2024mental,girgin2024sentiment}. Her bir problem kendine has çözümlere ihtiyaç duysa da bu çözümlerin temel DDİ işlevlerine olan ihtiyacı ortaktır.

Birçok DDİ uygulaması, özellikle yapay zeka ve makine öğrenmesi içerikli olanlar, tümcelerin ve sözcüklerin özel bir şekilde temsil edilmesine ihtiyaç duyar. Bu gösterim yöntemlerinden biri Sözcük Gömme'dir. Gömme yönteminde sözcükler, tamlamalar ve tümceler matematiksel vektörlere dönüştürülür. Dilbilimsel yapıların birbirleriyle olan anlamsal ve sözdizimsel özelliklerini içeren bu vektörler, DDİ problemlerinin çözümlerinde önemli bir yere sahiptir. Öyle ki, gömme yönteminin performansı, ilgili sistemin başarısını doğrudan etkiler.

Türkçe çevrimiçi içeriğin hızlıca artması, dilin zengin morfolojisi ve anlamsal incelikleriyle başa çıkabilen sağlam geri getirme sistemlerine olan gereksinimi artırmıştır. Ancak geleneksel gömme yöntemleri, kaynak ve veri kümesi çeşitliliğinden İngilizceye odaklı olarak geliştirilmiştir. Türkçe gibi, kaynak bakımından çok da zengin olmayan diller, kendine özgü modellerin azlığından muzdariptir. Ayrıca, çok dilli modellerin dil bazlı başarıları veya makine tarafından çevrilmiş veri üzerine kurulan modellerin başarıları yeterli olmayabilir. Bu gibi kısıtlar, sözcük gömmenin önemli olduğu geri getirme sistemlerinin performansını olumsuz etkileyebilir.

Türkçe DDİ alanında, dilin morfolojik karmaşıklığını etkin bir şekilde ele alan birkaç model önemli ilerlemeler kaydetmiştir. Öne çıkan örnekler arasında, kelime türü etiketleme (POS) ve adlandırılmış varlık tanıma (NER) gibi görevlerde güçlü performans sergileyen BERTurk \cite{BERTurk}, BertT5urk \cite{BERTurkv2}, Loodos Türkçe BERT \cite{loodos_bert} ve VNLP \cite{turker2024vnlpturkishnlppackage} yer almaktadır. Geri getirme görevleri için ise ColBERT \cite{colBERT-architecture} ve Turkish-colBERT \cite{kesgin2023developing} gibi modeller, sorgu-belge ilişkilerini modellemede başarılı olmuştur. Ancak, özellikle Türkçe gibi dillerde bu modellerin performansı, dil özelinde ön işlemenin kritik rolünü vurgulayan çalışmalarla daha da geliştirilebilir \cite{arslan2008comparison,yilmazel2010language}.

Bu çalışmada, Türkçe için geri getirme görevleri ile ilgili problemleri giderebilecek ve diğer alanlar için de ince-ayar yapılarak güçlü bir temel model sağlayacak olan bir gömme modeli olan TurkEmbed4Retrieval'ı sunuyoruz.

Bu çalışmanın başlıca katkıları şunlardır:
\begin{itemize}
    \item Türkçe'ye özel olarak geliştirilmiş bir gömme modelinin geri getirme görevlerine adaptasyonu.
    \item Matruşka temsili öğrenme tekniğinin kullanılmasıyla daha esnek ve çok boyutlu gömmelerin oluşturulması.
    \item Modelimizin, mevcut en iyi model olan Turkish-colBERT'ten daha iyi performans göstermesi.
\end{itemize}

Makalenin geri kalanı şu şekilde düzenlenmiştir: \ref{literature}. bölümünde ilgili çalışmaları, \ref{methodology}. bölümünde modelin mimarisi ve ince ayar süreci detaylandırılmaktadır. \ref{experiments}. bölümünde test ortam, veri kümeleri ve karşılaştırmalı değerlendirme tanımlanmaktadır. Bölüm \ref{results}, sonuçları değerlendirir ve elde edilen performans iyileştirmelerini analiz eder.
Son olarak, bölüm \ref{conclusion}, çalışmayı özetler ve gelecekteki yönelimleri tartışır.

\section{İlg{\footnotesize İ}l{\footnotesize İ} Çalışmalar}\label{literature}
Kelime gömmeleri, kelimelerin anlamsal ve sözdizimsel özelliklerini yakalayan vektör temsilleridir ve doğal dil işleme (DDİ) uygulamaları için temel bir öneme sahiptir. Word2Vec \cite{word2vec}, GloVe \cite{glove} ve FastText \cite{fasttext} gibi erken dönem statik modeller, kelimeler için bağlamdan bağımsız sabit vektörler sunar. Bu modeller, özellikle Türkçe gibi morfolojik olarak zengin dillerde bağlama dayalı anlamları yakalamakta zorlanır.

ELMo \cite{elmo} ve BERT \cite{bert} gibi bağlamsal modeller, bağlama göre değişen dinamik gömmeler sunarak karmaşık görevlerde performansı önemli ölçüde artırmıştır. Türkçe için bu modeller, eklerin kelime anlamlarını veya işlevlerini değiştirebildiği birleştirici (agglutinative) yapısı nedeniyle kritik öneme sahiptir. Statik ve bağlamsal modellerin karşılaştırıldığı bir çalışma \cite{Sar_ta__2024}, BERT gibi bağlamsal modellerin hem içsel hem de dışsal görevlerde statik modellerden üstün olduğunu ortaya koymuştur.

Türkçe DDİ, dilin morfolojik karmaşıklığını ele alma ihtiyacıyla önemli ilerlemeler kaydetmiştir. 
35 GB'lik bir korpus üzerinde (Türkçe Vikipedi ve OSCAR dahil) eğitilen BERTurk\cite{BERTurk} ve en son yayınlanan FineWeb2 veri kümesinin Türkçe kısmı ile ön-eğitim yapılan 1.42 milyar parametreli ve T5 mimarili BertT5urk \cite{BERTurkv2} , Türkçe DDİ için bir ölçüt olup morfolojiyi etkin bir şekilde işler. Modellerin bu çalışmalarda, POS etiketleme ve Adlandırılmış Varlık Tanıma (NER) gibi görevlerde güçlü performans gösterdiği detaylandırılmıştır. "Loodos/Türkçe Dil Modelleri" \cite{loodos_bert} deposu Türkçe için ince ayar yapılmış olup çeşitli DDİ görevleri için kaynak sunar. Türkçe DDİ için kapsamlı bir paket olan VNLP \cite{turker2024vnlpturkishnlppackage}, önceden eğitilmiş gömmeler ve duygu analizi ile NER için araçlar içerir. VNLP, açık kaynak yapısını ve kullanım kolaylığını vurgular; önceden eğitilmiş gömmeler büyük Türkçe korpuslar üzerinde geliştirilmiştir. Geri getirme alanında, ColBERT \cite{colBERT-architecture} gibi modeller, geç etkileşim (late interaction) mekanizmalarıyla sorgu-belge ilişkilerini başarıyla modellemiştir. Türkçe için Turkish-colBERT \cite{kesgin2023developing}, MS-Marco \cite{ms-marco} veri setinin 500 bin örneği üzerinde eğitilmiş bir model olarak öne çıkar.

Bu çalışma, modelin Türkçe kapasitesinin All-NLI-TR \cite{budur-etal-2020-data} ve STSb-TR \cite{beken-fikri-etal-2021-semantic} veri kümelerinde eğitim ile artırılarak doğal dil çıkarımı ve tümce metin benzerliği görevlerinin iyileştirilmesini sağlamıştır. Bu eğitim sonrasında ise MS-Marco-TR veri seti ve gelişmiş tekniklerin uygulanması ile daha önce bahsedilen çalışmaların performansı aşılmıştır.

\section{Metodoloj{\footnotesize İ}}\label{methodology}
\begin{figure}[ht!]
    \centering
    \shorthandoff{=}
    \includegraphics[width=0.75\linewidth]{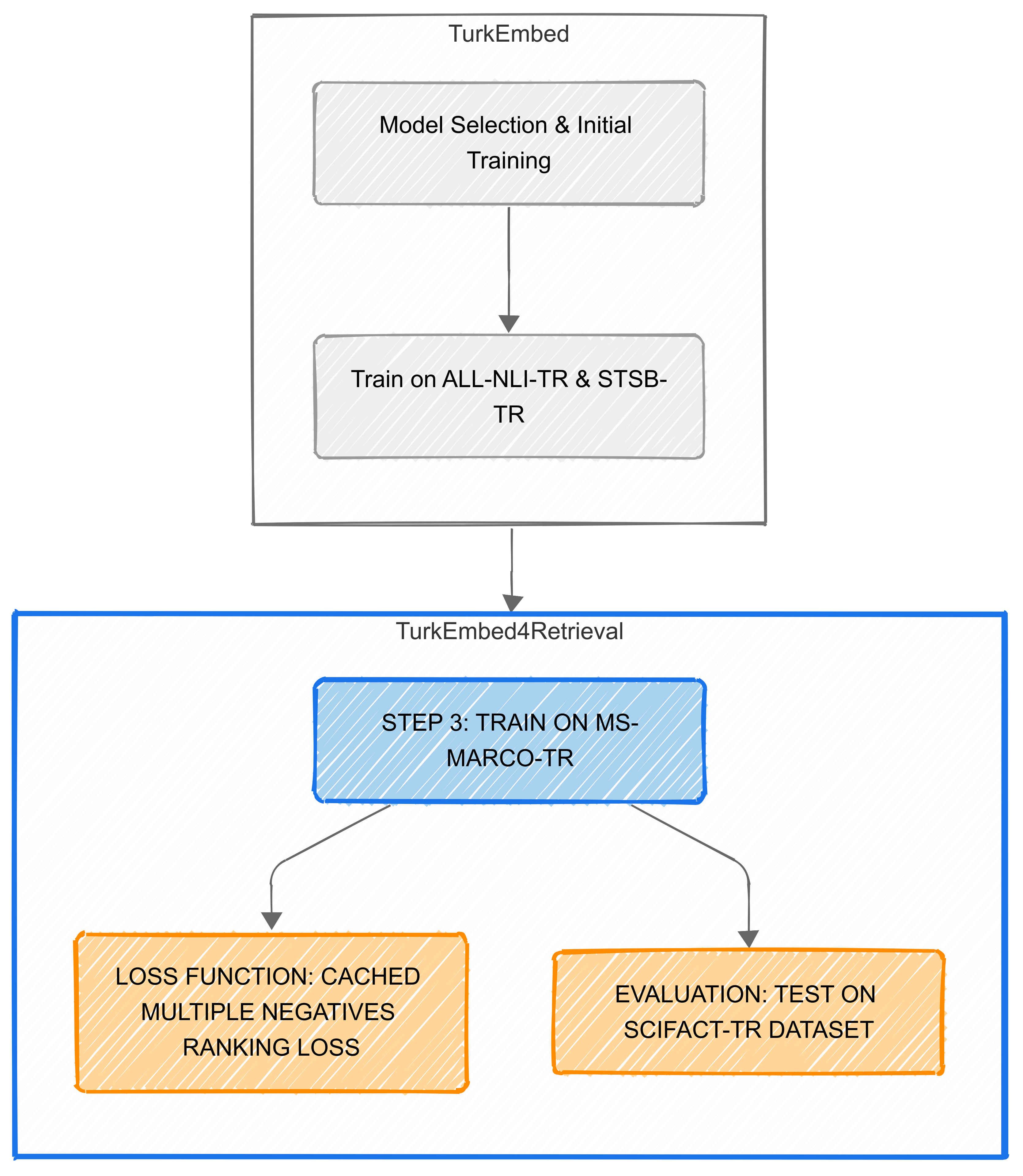}
    \shorthandon{=}
    \caption{TurkEmbed4Retrieval Sıralı Eğitim Veri Hattı}
    \label{fig:turkembedflowchart}
\end{figure}

TurkEmbed4Retrieval modelinin geliştirilmesi, Şekil \ref{fig:turkembedflowchart}’de detaylı bir akış şeması olarak gösterilen çok aşamalı bir metodolojiyi takip eder. Bu süreç, modelin güçlü bir temel üzerine inşa edilmesini, Türkçe’ye özgü veri kümeleriyle ardışık olarak eğitilmesini ve nihayetinde bilgi erişim görevleri için optimize edilmesini sağlamaktadır.

\subsection{Aşama 1: Model Seçimi}
Geliştirme sürecinin ilk aşamasında, çok sayıda önceden eğitilmiş model değerlendirilmiş ve başlangıç benchmarkları gerçekleştirilerek en uygun temel model seçilmiştir. Bu değerlendirmeler sonucunda, 305 milyon parametreye sahip GTE-multilingual-base modeli \cite{zhang2024mgte}, üstün performansı nedeniyle tercih edilmiştir. GTE-multilingual-base, Matruşka Temsili Öğrenme (Matryoshka Representation Learning) \cite{kusupati2022matryoshka} tekniğini kullanarak çok boyutlu gömmeler üretir. Bu gömmeler, Türkçe’nin morfolojik karmaşıklığını etkili bir şekilde temsil etme kapasitesiyle, modelin Türkçe DDİ görevleri için uygunluğunu artırmaktadır.

\subsection{Aşama 2: Ardışık Eğitim ve Değerlendirme Süreci}
Modelin Türkçe dil anlama yetkinliğini geliştirmek amacıyla, ikinci aşamada iki temel veri kümesi üzerinde ardışık eğitim uygulanmıştır.

\subsubsection{ALL-NLI-TR Üzerinde Eğitim}
Model, Türkçe Doğal Dil Çıkarımı (DDÇ) için tasarlanmış ALL-NLI-TR veri kümesi üzerinde eğitilmiştir. Bu veri kümesi, 482.091 eğitim örneği içermektedir ve modelin anlamsal olarak ilişkili ve ilişkisiz cümle çiftlerini ayırt etme yeteneğini optimize etmek için Multiple Negatives Ranking Loss fonksiyonu kullanılmıştır. Eğitim sonrasında, modelin performansı ALL-NLI-TR doğrulama seti üzerinde değerlendirilmiştir.

\subsubsection{STSB-TR Üzerinde Eğitim}
Bir sonraki aşamada, model, Türkçe Anlamsal Metin Benzerliği (AMB) için 5.749 örnek içeren STSB-TR veri kümesi üzerinde eğitilmiştir. Bu aşamada, cümleler arasındaki benzerlik ölçümünü iyileştirmek için CoSENT Loss fonksiyonu uygulanmıştır. Eğitim tamamlandıktan sonra, modelin performansı STSB-TR kıyaslama kümesi üzerinde test edilmiştir.

Bu ardışık eğitim ve değerlendirme süreci, Türkçe DDÇ ve AMB görevleri için özel olarak tasarlanmış TurkEmbed modelinin oluşturulmasını sağlamıştır.

\subsection{Aşama 3: Son Eğitim ve Değerlendirme}
Bilgi erişim görevlerine yönelik olarak, TurkEmbed modeli üçüncü aşamada MS-Marco-TR \cite{msmarco-tr}  veri kümesi üzerinde ince ayar yapılarak TurkEmbed4Retrieval haline getirilmiştir. MS-Marco-TR, MS-Marco veri kümesinin Türkçeye çevrilmiş bir versiyonu olup, her biri bir olumlu ve bir olumsuz metinle eşleştirilmiş 1 milyon sorgu-metin çifti içermektedir. İnce ayar sırasında, modelin ilgili metinleri sıralamada daha başarılı olmasını sağlamak için Cached Multiple Negatives Ranking \cite{gao2021scalingdeepcontrastivelearning} kayıp fonksiyonu kullanılmıştır. Bu süreç, modelin bilgi erişim performansını optimize etmiştir.

Son olarak, TurkEmbed4Retrieval’in etkinliği, SciFact-TR veri kümesi üzerinde test edilerek değerlendirilmiştir. Bu değerlendirme, modelin Türkçe sorgular için ilgili belgeleri geri getirme yeteneğini ölçmüştür.

\section{Testler}\label{experiments}

\subsection{Test Ortamı}

Modelin eğitimi ve testleri için yüksek performanslı NVIDIA A100 model 40 GB'lik hafızaya sahip grafik kartı içeren bir sistem kullanılmıştır. Sistemin yazılımı Python 3.11.11 üzerinden PyTorch 2.5.1+Cu121, Sentence Transformers 3.3.1, Transformers 4.49.0.dev0, Accelerate 1.2.1, Datasets 3.2.0 ve Tokenizers 0.21.0 kütüphanelerini içeren bir ortam olarak hazırlanmıştır. 

\subsection{Ver{\footnotesize İ} Kümes{\footnotesize İ} ve Ölçütler}

Modelimizin eğitimi için MS-Marco\cite{ms-marco} veri kümesinin Türkçeye çevrilmiş hali olan MS-Marco-TR\cite{msmarco-tr}'yi kullandık. Bu veri seti, her biri belirli bir sorguyla ilgili bir pozitif ve bir negatif metinle eşleştirilen 1 milyon örnekten oluşmaktadır. 

Test için SciFact\cite{sci-fact} veri setinin Türkçeye çevrilmiş hali SciFact-TR'yi\cite{scifact-tr} kullandık. Bu set 6292 örneklem içerir. Örneklemler, sorguların dökümanlarla eşleştirilmesinden oluşur. Her bir döküman, bir başlık ve bir metin içerir.

Bilgi erişim sistemimizi değerlendirmek için Duyarlılık@1, 5 ve 10, MRR@10 ve NDCG@10 ölçütlerini kullanıyoruz. Duyarlılık ölçütünde ilk 1, 5 ve 10 benzerlik sonucunda ilgili belgelerin ne kadarını geri getirdiğimizi ölçer. MRR@10, ilk doğru sonucun ilk 10 sonuç içindeki sıralamasını değerlendirerek hızı analiz eder. NDCG@10 ise ilk 10 sonucun sıralama kalitesini ölçer ve ilgili belgelerin üst sıralarda yer almasını ödüllendirir. Bu ölçütler, sistemimizin sıralama, hız ve kapsamlılık performansını dengeli bir şekilde değerlendirmemizi sağlar.

\subsection{Karşılaştırmalı Değerlendirme}
TurkEmbed4Retrieval ile Turkish-colBERT modellerinin geri getirme performansları Tablo \ref{table:1}'de karşılaştırmalı olarak verilmiştir.

\begin{table}[ht!]
    \centering
    \caption{\textsc{Ger{\footnotesize İ} Get{\footnotesize İ}rme Performansı Kıyaslaması}}
    \label{table:1}
    \begin{tabular}{|l|l|l|l|}
        \hline
        \textbf{Ölçüt} & \textbf{TurkEmbed4Retrieval} & \textbf{Turkish‐} & \textbf{Göreceli } \\
                       &                              & \textbf{colBERT} & \textbf{İyileşme (\%)}\\
        \hline
        Duyarlılık @1 & 0.7116 & 0.4838 & +19.26\% \\
        \hline
        Duyarlılık @5 & 0.9003 & 0.6785 & +32.68\% \\
        \hline
        Duyarlılık @10 & 0.9417 & 0.7552 & +24.73\% \\
        \hline
        MRR@10 & 0.8221 & 0.5688 & +28.25\% \\
        \hline
        NDCG@10 & 0.8484 & - &  -  \\
        \hline
    \end{tabular}
\end{table}

\begin{table*}[!h]
    \centering
    \caption{\textsc{MS-Marco-TR Ver{\footnotesize İ} Kümes{\footnotesize İ} Üzer{\footnotesize İ}nden Yapılan İnce Ayarın Model Performansına Etk{\footnotesize İ}s{\footnotesize İ}}}
    \label{table:2}
    \begin{tabular}{|c|c|c|c|c|c|}
        \hline
        \multirow{2}{*}{\textbf{Ölçüt}} & \multicolumn{5}{c|}{\textbf{İnce Ayar Öncesi $\rightarrow$ İnce Ayar Sonrası}} \\
        \cline{2-6}
                               & \textbf{@1} & \textbf{@3} & \textbf{@5} & \textbf{@10} & \textbf{@100} \\
        \hline
        \textbf{Doğruluk} & 0.7400 $\rightarrow$ 0.7533 & 0.8300 $\rightarrow$ 0.8767 & 0.8700 $\rightarrow$ 0.9067 & 0.8967 $\rightarrow$ 0.9433 & - \\
        \hline
        \textbf{Kesinlik} & 0.7400 $\rightarrow$ 0.7533 & 0.3067 $\rightarrow$ 0.3200 & 0.1960 $\rightarrow$ 0.2040 & 0.1017 $\rightarrow$ 0.1070 & - \\
        \hline
        \textbf{Duyarlılık} & 0.6983 $\rightarrow$ 0.7116 & 0.8197 $\rightarrow$ 0.8619 & 0.8640 $\rightarrow$ 0.9003 & 0.8937 $\rightarrow$ 0.9417 & - \\
        \hline
        \textbf{NDGC@10} & - & - & - & 0.8169 $\rightarrow$ 0.8484 & - \\
        \hline
        \textbf{MRR@10} & - & - & - & 0.7957 $\rightarrow$ 0.8221 & - \\
        \hline
        \textbf{MAP@100} & - & - & - & - & 0.7937 $\rightarrow$ 0.8188 \\
        \hline
    \end{tabular}
\end{table*}

\section{Sonuçlar ve Tartışma}\label{results}
Yapılan karşılaştırmalı değerlendirme, TurkEmbed4Retrieval modelinin Turkish-colBERT’e kıyasla tüm ölçütlerde üstün performans sergilediğini göstermektedir. Duyarlılık @1’de \%19.26, duyarlılık @5’te \%32.68 ve duyarlılık @10’da \%24.73’lük göreceli iyileşmeler elde edilmiştir. Ayrıca, MRR@10 ölçütünde de \%28.25’lik artış gözlemlenmiştir. NDCG@10 metriğinde ise Turkish-colBERT modeli için bir başarı değeri paylaşılmamıştır.

Bu başarının temel nedenleri arasında, MS-Marco-TR veri seti üzerinde yapılan ince ayar ve matruşka temsili öğrenme tekniğinin sağladığı esneklik sıralanabilir. İnce ayar, modelin Türkçe sorgu-kesit ilişkilerini daha iyi anlamasını sağlamış, özel kayıp fonksiyonu ise negatif örneklerin sıralanmasında etkin bir optimizasyon sunmuştur. İnce-ayar yapılan modelimizin performansına pozitif etki eden optimizasyonlardan biri de küme büyüklüğünü olabildiğince büyük seçmektir çünkü yapılan deneylerde Cached Multiple Negatives Ranking Loss'un 4096 küme büyüklüğüne kadar performans artışı sağladığı görülmüştür \cite{gao2021scalingdeepcontrastivelearning}. Biz de bu kayıp fonksiyonu sayesinde sahip olduğumuz donanımın kapasitesini olabildiğince kullanarak küme büyüklüğünü 1024'e kadar artırıp yaşadığımız limitasyonu mümkün mertebe azaltmaya çalıştık. 

Tablo \ref{table:2}, ince ayarın model performansını önemli ölçüde artırdığını doğrulamaktadır. Özellikle duyarlılık @10’da \%4.8’lik bir iyileşme, modelin geri getirme doğruluğunu büyük ölçüde geliştirdiğini göstermektedir. Bu sonuçlar, Türkçe gibi morfolojik olarak karmaşık dillerde görev-özel eğitimlerin kritik önemini vurgular.

Tablo \ref{table:3} ise TurkEmbed4Retrieval modelinin geri getirme (retrieval) görevi için MS-Marco-TR veri kümesi ile eğitildikten sonra anlamsal metin benzerliği (AMB) görevindeki performansındaki değişimi göstermektedir. Bu tablodaki değerler bizlere TurkEmbed4Retrieval modelinin aynı anda geri getirme görevi için en üstün performansı sağlarken, yapılan optimizasyonlar sayesinde önceki eğitilmiş görevleri de en az derecede unuttuğunu göstermektedir.  

\begin{table}[ht!]
    \centering
    \caption{\textsc{Model{\footnotesize İ}n Ger{\footnotesize İ} Get{\footnotesize İ}rme görev{\footnotesize İ} İç{\footnotesize İ}n İnce-ayarlanmasından Sonra Anlamsal Met{\footnotesize İ}n Benzerl{\footnotesize İ}ğ{\footnotesize İ} Görevindek{\footnotesize İ} Başarımının Değ{\footnotesize İ}ş{\footnotesize İ}m{\footnotesize İ}}}
    \label{table:3}
    \begin{tabular}{|c|c|c|}
        \hline
        \textbf{Benzerlik} & \multicolumn{2}{c|}{\textbf{İnce Ayar Öncesi $\rightarrow$ İnce Ayar Sonrası}} \\
        \cline{2-3}
        \textbf{Ölçütü}     & \textbf{Pearson (r)} & \textbf{Spearman (p)}  \\
        \hline
        \textbf{Kosinüs} & 0.8455 $\rightarrow$ 0.7896 & 0.8534 $\rightarrow$ 0.7773  \\
        \hline
        \textbf{Skaler Çarpım} & 0.8455 $\rightarrow$ 0.7896 & 0.8534 $\rightarrow$ 0.7773  \\
        \hline
        \textbf{Öklidyen} & 0.8461 $\rightarrow$ 0.7868 & 0.8534 $\rightarrow$ 0.7773  \\
        \hline
        \textbf{Manhattan} & 0.8453 $\rightarrow$ 0.7896 & 0.8527 $\rightarrow$ 0.7773  \\
        \hline
    \end{tabular}
\end{table}

\section{Sonuç}\label{conclusion}

Bu çalışmada Türkçe için geri getirme görevlerinde kullanılmak üzere geliştirdiğimiz TurkEmbed4Retrieval modelini sunuyoruz.
TurkEmbed modelini MS-Marco-TR veri kümesi üzerinde ince ayar yaparak elde ettik. Test aşamasında, modelin başarısını ölçmek için SciFact-TR veri kümesini kullandık. Testler, modelimizin Turkish-colBERT'i geride bıraktığını ve daha iyi performans sergilediğini göstermiştir. 

TurkEmbed4Retrieval, Türkçe DDİ alanında bilgi geri getirme sistemleri açısından önemli bir adım olup matruşka temsili öğrenme yönteminin ve büyük ölçekli veri kümelerinin kullanılmasının önemini ortaya koymuştur. Bu çalışmada farklı alanlarda kullanılabilecek temel bir model sunuyoruz. 
İleriki çalışmalarımızda, modelin soru-cevap, belge sıralama ve benzeri farklı görevler için özelleştirilmesini planlıyoruz. Buna ek olarak, doğal ve sentetik verilerin harmanlanmasıyla oluşturulacak daha büyük veri kümelerini kullanarak temel modelin kalitesini artırmayı amaçlıyoruz. Yine sentetik veri desteği ile alanlara özel üretilmiş veri kümeleri oluşturarak, model için alan odaklı ince ayar eğitimleri gerçekleştirmeyi planlıyoruz.

\end{document}